\documentclass[12pt,preprint]{emulateapj}

\shorttitle{The progenitor mass of SGR1900+14}
\shortauthors{Davies et al.}
\usepackage{graphicx}

\usepackage{lscape}
\usepackage[bf,small]{caption}

\newcommand{\microns}{$\mu$m}

\def\hei{He\,{\sc i}}
\def\heii{He\,{\sc ii}}
\def\civ{C\,{\sc iv}}
\def\niii{N\,{\sc iii}}

\def\ga{\mathrel{\hbox{\rlap{\hbox{\lower4pt\hbox{$\sim$}}}\hbox{$>$}}}}
\def\la{\mathrel{\hbox{\rlap{\hbox{\lower4pt\hbox{$\sim$}}}\hbox{$<$}}}}

\def\msun{$M$\mbox{$_{\normalsize\odot}$}}

\def\lsun{$L$\mbox{$_{\normalsize\odot}$}}
\def\kms{\,km~s$^{-1}$}

\def\arcsec{$^{\prime \prime}$}

\begin{document}
\title{The progenitor mass of the magnetar SGR1900+14}
\author{Ben Davies\altaffilmark{1,2}, Don
  F.\ Figer\altaffilmark{2}, Rolf-Peter Kudritzki\altaffilmark{3},
  Christine Trombley\altaffilmark{2}, \\ Chryssa
  Kouveliotou\altaffilmark{4}, Stefanie Wachter\altaffilmark{5} }

\affil{$^{1}$School of Physics \& Astronomy, University of Leeds,
  Woodhouse Lane, Leeds LS2 9JT, UK.}

\affil{$^{2}$Chester F.\ Carlson Center for Imaging Science, Rochester
Institute of Technology, 54 Lomb Memorial Drive, Rochester NY, 14623,
USA} 

\affil{$^{3}$Institute for Astronomy, University of Hawaii, 2680
Woodlawn Drive, Honolulu, HI, 96822, USA} 

\affil{$^{4}$Space Science Office, VP62, NASA/Marshall Space Flight Center, Huntsville AL 35812, USA}

\affil{$^{5}$Spitzer Science Center, 1200 E California Blvd,
  California Institute of Technology, Pasadena, California 91125, USA}

\begin{abstract}
  Magnetars are young neutron stars with extreme magnetic fields
  (B$\ga$10$^{14}$-10$^{15}$\,G). How these fields relate to the
  properties of their progenitor stars is not yet clearly
  established. However, from the few objects associated with young
  clusters it has been possible to estimate the initial masses of the
  progenitors, with results indicating that a very massive progenitor
  star ($M_{\rm prog} >$40\msun) is required to produce a magnetar.
  Here we present adaptive-optics assisted Keck/NIRC2 imaging and
  Keck/NIRSPEC spectroscopy of the cluster associated with the
  magnetar SGR~1900+14, and report that the initial progenitor star
  mass of the magnetar was a factor of two lower than this limit,
  $M_{\rm prog}$=17$\pm$2\msun. Our result presents a strong challenge
  to the concept that magnetars can only result from very massive
  progenitors. Instead, we favour a mechanism which is dependent on
  more than just initial stellar mass for the production of these
  extreme magnetic fields, such as the ``fossil-field'' model or a
  process involving close binary evolution.
\end{abstract}

\keywords{open clusters \& associations: individual (Cl~1900+14),
  stars: evolution,
  stars: neutron, stars: individual (SGR1900+14)
}


\section{Introduction} \label{sec:intro}
Magnetars are currently recognized as a distinct group of neutron
stars comprising of several classes of object, such as Soft Gamma
Repeaters (SGRs), Anomalous X-ray Pulsars (AXPs), and some Compact
Central Objects. These objects are characterized by relatively long
spin periods and large spin-down torques, which imply magnetic fields
on the order of B$\ga10^{14}-10^{15}$\,G, making them the most highly
magnetized objects known in the Universe
\citep{D-T92,T-D95,Kouveliotou98,Mereghetti08}. They are also known to
enter active periods during which they emit very intense ($10^{37}\la
L \la10^{41}$ erg/s), short ($\sim$0.1\,s) hard X-/gamma ray bursts,
as well as extremely energetic Giant Flares of $\ga10^{44-46}$ ergs
lasting several minutes.

It is still unclear how magnetars are formed. The current theoretical
framework for magnetar production requires that the core of a massive
star has a very fast rotation speed in the first few seconds after it
goes supernova (SN). If the rotation period is shorter than the
convective timescale within the neutron star -- about 1ms -- a
highly-efficient dynamo operates which boosts the magnetic field to
$\sim$1000 times that of a `regular' neutron star, and very rapidly
slows the rotation period down to of order 1 second in a matter of
years \citep[the so-called `dynamo'
  mechanism,][]{D-T92,TCQ04}. However, recent stellar evolution
calculations show that the cores of massive stars are substantially
spun down as they enter the Red Supergiant (RSG) phase through
magnetic braking between the stellar core and convective envelope
\citep{Heger05}. Thus, the problem exists of how the core of a massive
star can retain sufficient angular momentum through to the SN stage
such that the post-SN core is able to jump-start the dynamo
mechanism. It has been suggested that those stars with $M_{\rm init}
\ga $40\msun\ are able to lose a substantial fraction of their
hydrogen-rich envelope while still on the main-sequence, allowing them
to skip the RSG phase, and therefore avoid the severe spin-down of the
core as the outer envelope expands and becomes convective
\citep{Gaensler05apj}.

Where magnetars have been associated with star-clusters it has been
possible to estimate the initial mass of the progenitor
empirically. Evidence suggests that the magnetar phase is short and
that the SN that produced it must have occured recently
\citep[$\la10^4$yrs ago,][]{Kouveliotou94}, such that the age of the
cluster is much greater than the lifetime the magnetar. Consequently,
by measuring the age of the star-cluster we can determine the age of
the progenitor star when it went SN. Then, as a star's lifetime is a
strong function of its initial mass, we can estimate the initial mass
of the magnetar's progenitor. In the cases of the magnetars
SGR~1806-20 and CXOU~J164710.2-455216, associated with the clusters
Cl~$1806-20$ and Westerlund~1 (Wd~1) respectively, it appears that the
magnetar progenitors had initial masses
$\ga$40\msun\ \citep{Figer05sgr,Bibby08,Muno06}. These results are
therefore consistent with the hypothesis that magnetars decend from
the most massive stars. Further supporting evidence for this
hypothesis comes from a study of the expanding H{\sc i} shell around
the magnetar 1E~1048.1-5937. When the shell was interpreted as a
stellar wind bubble blown by the progenitor, a progenitor mass of
30-40\msun\ was inferred \citep{Gaensler05}.

There is a fourth magnetar, SGR~$1900+14$, which can be used to test
this hypothesis. It too is thought to belong to a cluster, which was
first recognized for its two bright Red Supergiant (RSG) members
\citep{Vrba96,Vrba00}. However, up until now this cluster has been
poorly studied. The only current distance estimate is a
spectrophotometric distance which assumes an intrinsic brightness for
the RSGs, and so cannot be used to derive accurate luminosities for
the RSGs themselves. Consequently, any estimate for the magnetar's
progenitor mass based on the RSG luminosities is unreliable. The best
evidence for the association of the magnetar and star-cluster comes
from the detection of an infrared ring around the source, analysis of
which placed the magnetar at the same distance from Earth as the
cluster's spectrophotometric distance \citep{Wachter08}.
 
In this paper we present a spectroscopic and photometric analysis of
the stellar content of the cluster Cl~1900+14. We determine a
kinematic distance to the cluster and derive the cluster's
age. Ultimately, we are able to establish an estimate for the initial
mass of the progenitor of SGR~1900+14. We begin in
Sect.\ \ref{sec:obs} with a description of the observations and data
reduction procedure. In Sect.\ \ref{sec:res} we present the results of
our analysis, and in Sect.\ \ref{sec:disc} we discuss the implication
for the evolution of massive stars.


\section{Observations \& data reduction} \label{sec:obs}

\subsection{Spectroscopy}
Spectroscopic observations were made using NIRSPEC, the infra-red
spectrograph mounted on Keck-II, during the night of 23 Jun
2008. High-resolution observations were made of the stars A and B
\citep[as identified by][]{Vrba00} with the spectrograph in
cross-dispersed mode, the cross-disperser angle set to 35.53\degr\ and
the echelle angle to 62.53\degr. We used the NIRSPEC-7 filter and the
0.576\arcsec$\times$24\arcsec\ slit, which gave a spectral resolution
of R$\approx$20,000 in select regions of the K-band.

Low-resolution observations were also made of several other stars
within the field of the cluster. Here, the echelle was replaced with a
mirror, which when combined with the cross-disperser and the
42\arcsec$\times$0.57\arcsec\ slit gave a spectral resolution of
$R$=1000 in the wavelength range 1.9-2.4\microns.

Repeated observations were made of each target, with the star nodded
along the slit by $\ga$ 10\arcsec. In addition to the target stars, we
also observed the telluric standard stars HD179282 and HD173003 to
characterize the atmospheric absorption features in the
$K$-band. Continuum-lamp exposures were made for flat-field purposes,
arc lamps were observed for wavelength calibration, while in
high-resolution mode we also observed the continuum lamp through the
etalon filter to sample the gaps in between the arc lines.

Our data-reduction procedure included the subtraction of nod-pairs to
remove diffuse sky emission, and division by a normalized flat-field
frame to correct for variations in pixel sensitivity on the
detector. Before extraction, the spectral traces were resampled onto a
grid linear in both the spatial and dispersion directions using the
arc and etalon frames. This rectification process provides absolute
wavelength calibration accurate to $\pm$4\kms\ for the high-resolution
spectra -- a detailed description of this part of the data reduction
is given in \citet{Figer03}. Atmospheric absorption features were
removed by dividing through by a telluric standard star, which first
had its intrinsic spectral features removed using a synthesised
spectrum appropriate for its spectral type.

\begin{figure}[t]
  \centering
  \includegraphics[width=8.5cm]{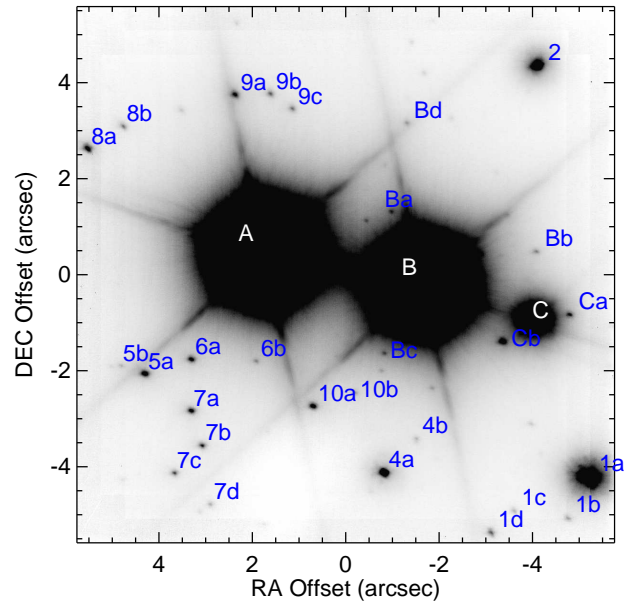}
  \caption{NIRC2 $K$-band image of the centre of the field, showing
    the fainter stars close to the two RSGs. We use the stellar
    identifications of \citet{Vrba00}; where we resolve one of Vrba et
    al.'s objects into multiple components, we sub-label them
    alphabetically. \\}
  \label{fig:hrfinder}
\end{figure}

\subsection{Imaging}
The field centered on the cluster was imaged using the Near Infra-Red
Camera 2 (NIRC2) on Keck-II, in combination with Laser Guide Star
Adaptive Optics (LGSAO), on 23 Aug 2008. Five overlapping 20 second
exposures were taken in the $H-$ and $K-$ bands, using the narrow
camera, in a X-shaped dither pattern. Raw images were cleaned by
subtraction of darks and flat fielding, then median combined.

Aperture photometry of point sources was done using IDL scripts,
including those in the IDL-adapted version of DAOPHOT.  Point-sources
which were too close to the two saturated bright stars in the centre
were rejected. Magnitudes were calibrated from a control frame taken
the same night, and by using 2MASS photometry of bright, isolated,
non-saturated stars. For astrometric calibration we used archival HST
observations of a field containing the cluster, and used stars 9a, 7a,
and 4 (see Fig.\ \ref{fig:hrfinder}) to compute the geometric
transformation.

\begin{figure}
  \centering
  \includegraphics[width=8.5cm]{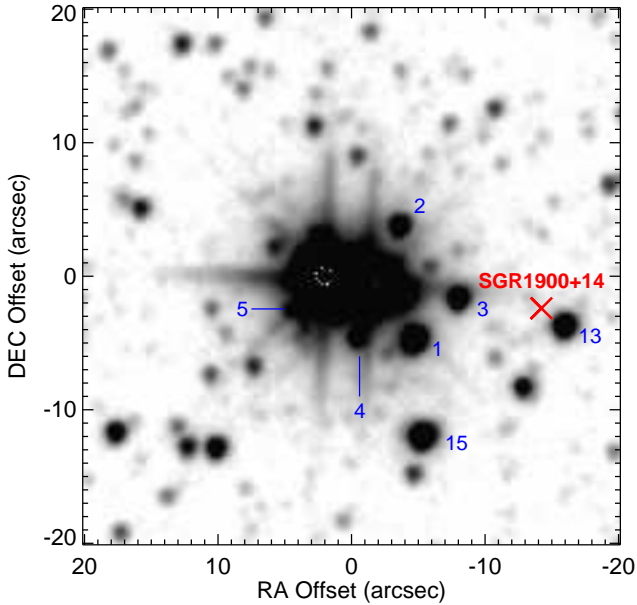}
  \caption{Wide-field UKIDSS $H$-band image of Cl~1900+14, indicating
    the stars of which low-resolution spectra were obtained. The
    position of the magnetar SGR1900+14 is indicated.}
  \label{fig:finder}
\end{figure}

In addition to the Keck/NIRC2 photometry, we also obtained photometry
of stars over a wider field from the UKIDSS survey of the Galactic
Plane \citep{Lucas08,Lawrence07,Casali07,Hewitt06,Hambly08}. In
Fig.\ \ref{fig:finder} we show this wide-field image, and indicate the
stars for which we obtained spectroscopy.

\section{Results \& analysis} \label{sec:res}

\begin{figure}
  \centering
  \includegraphics[width=8.5cm,bb=30 0 651 311]{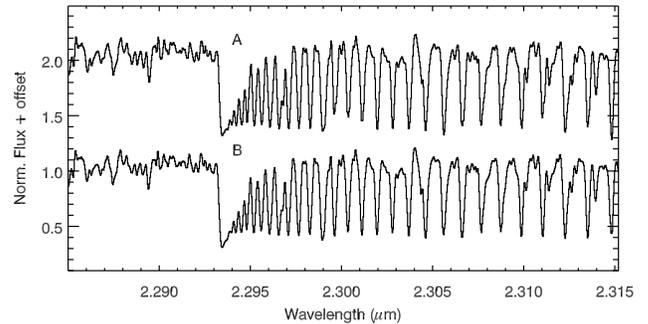}
  \caption{High-resolution spectra of the two RSGs in the region of
    the CO bandhead absorption feature.\\}
  \label{fig:rsgspec}
\end{figure}

The high-resolution $K$-band spectra of the two stars $A$ and $B$ in
the region of the CO bandhead are presented in
Fig.\ \ref{fig:rsgspec}, with both objects displaying the deep
absorption signature of luminous late-type stars. The star `C' was
shown by V96 to be a foreground object. To derive physical parameters
for the two stars, we follow the method of \citet{RSGC2paper}. To
determine a spectral type for each star we measured the equivalent
width of the CO bandhead and compared to similar measurements of
template stars. We found absorption strengths that were in excess of
that seen in giant stars, and were more typical of M1-2
supergiants. Below we describe our analysis procedure in more detail,
the results of which are presented in Table \ref{tab:rsg}.

\subsection{Extinction} \label{sec:rsgs}

We measured line-of-sight extinctions from the $J-K$ and $H-K$
colour-excesses of the two stars, using the photometry of
\citet{Vrba96}, intrinsic colours from \citet{Levesque05} appropriate
for the stars' spectral types, and the extinction law of
\citet{R-L85}. We found comparable extinctions for both stars from
each IR colour. The $H-K$ extinction for the two RSGs, $A_{K} = 1.44
\pm 0.06$, implies a visual extinction of $A_{V} = 12.9 \pm 0.5$ if we
assume $A_{K}/A_{V} = 0.112$ \citep{R-L85}. This is in excellent
agreement with the extinction derived by \citet{Kaplan02}, $A_{V} =
12.8 \pm 0.8$, determined from converting X-ray absorption to an
equivalent hydrogen column density. However, it is very different from
the extinction derived by \citet{Vrba96}, who found $A_{V} \simeq
19.2$. The reason that such a large extinction was inferred by these
authors is not clear \citep[see also][]{Wachter08}, however it may be
that the extinction law they used was optimized for optical
wavelengths while they used near-IR colour indices to determine the
reddening.

\begin{table*}
  \centering
  \small
    \caption{{Physical properties of the Red Supergiants in
      Cl~1900+14.} \vspace{2mm}}
    \begin{tabular}{ccccccc}
      \hline \hline
      Star & v$_{\rm LSR}$ & $T_{\rm eff}$ & Spec type & $A_{\rm K}$ &
      $M_{\rm K}$ & log~($L$/\lsun) \\
       & (\kms) & (K) & ($\pm$2 subtypes) & & & \\
      \hline
      A & -14$\pm$4 & 3660$\pm$130 & M2 & 1.43$\pm$0.06 &
      -10.64$^{+0.20}_{-0.19}$ &   5.05$^{+0.08}_{-0.08}$ \\ 
      B & -17$\pm4$ & 3750$\pm$120 & M1 & 1.45$\pm$0.03 &
      -10.23$^{+0.18}_{-0.17}$ &   4.91$^{+0.07}_{-0.07}$ \\
      \hline \\
    \end{tabular}
  \label{tab:rsg}
\end{table*}


\subsection{Kinematic distance}
We measured radial velocities for the two stars by cross-correlating
the spectra with that of Arcturus presented in \citet{W-H96arct},
which had been degraded to the same spectral resolution as our
observations and shifted to the rest-frame. As with extinction, the
line-of-sight velocities we measure for the two RSGs are comparable to
within the errors. This is strong evidence that the objects are part
of a physical association. The mean velocity we measure\footnote{The
  uncertainty on the velocity is dominated by the absolute precision
  of our measurements.}, -15.5$\pm$4\kms, indicates a kinematic
distance of 12.5$\pm$0.3kpc using the Galactic rotation curve of
\citet{B-B93}. As we measure a negative velocity, there is no nearside
solution to the rotation curve along this line-of-sight. Though random
velocities which are peculiar to the Galactic rotation curve by up to
20\kms\ may not be uncommon, this would require the cluster to be
within 1kpc, which given the extinction to the object is highly
unlikely. If we take such a random motion as the dominant source of
uncertainty in the kinematic distance, we find 12.5$\pm$1.7kpc. This
distance is in good agreement with the spectrophotometric estimate by
\citet{Vrba96}.

\begin{figure*}
  \centering
  \includegraphics[width=12cm]{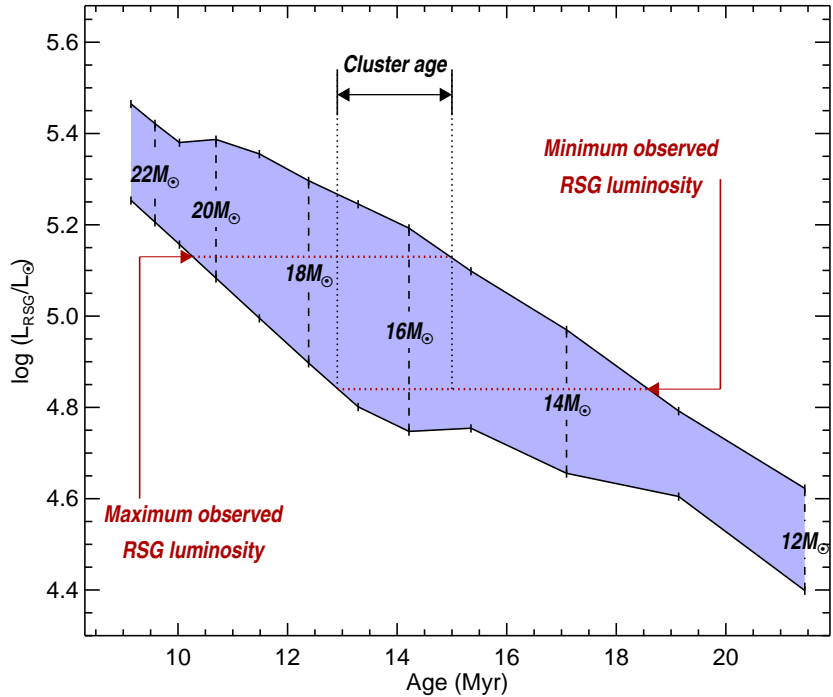}
  \caption{The minimum and maximum luminosities of Red Supergiants
    (RSGs) in a coeval cluster, as a function of cluster age,
    calculated using the Geneva rotating models at Solar
    metallicity\cite{Mey-Mae00}. The initial masses of the stars in
    the RSG phase are labelled. The red arrows
    indicate the range of RSG luminosities we observe in Cl~1900+14,
    $\pm 1 \sigma$. \\}
  \label{fig:minmax}
\end{figure*}

\subsection{Cluster age}
The distance, average extinction and the bolometric corrections of
\citet{Levesque05} were used to calculate intrinsic luminosities of
the RSGs in the cluster. As a cluster ages, the stars currently
experiencing the RSG phase will have decended from stars with lower
initial masses, and so will have lower luminosities. Therefore, from
the luminosity range of the RSGs in a cluster it is possible to
estimate the cluster's age.

Figure \ref{fig:minmax} shows the minimum and maximum luminosities of
Red Supergiants (RSGs) in a coeval cluster, as a function of cluster
age, calculated using the Geneva rotating models at Solar metallicity
\citep{Mey-Mae00}. The initial masses of the stars in the RSG phase
are labelled. The red arrows indicate the range of RSG luminosities we
observe in Cl~1900+14, $\pm 1 \sigma$. We can say that the cluster
{\it cannot} be younger than $\approx$13Myr, as the least luminous RSG
in such a cluster would be brighter than the faintest RSG in
Cl~1900+14. Similarly, we can place an upper limit to Cl~1900+14's age
of $\approx$15Myr, as the RSGs in a cluster older than this would be
fainter than the brightest RSG in Cl~1900+14. Therefore, from the RSG
luminosity range we observe in the cluster, we can constrain the age
to 14$\pm$1\,Myr. If a greater number of RSGs were to be present in
this cluster, we would expect them to occupy the luminosity range
indicated by the shaded region between 13-15Myr ($4.75 <
\log(L/$\lsun$) < 5.25$). The effect of using different evolutionary
models is discussed in Sect.\ \ref{sec:disc}.

\subsection{Cluster coevality}

We can assess the coevality of the cluster by studying its stellar
population. From our low-resolution spectroscopy of stars in the
cluster field, we separate the massive stars belonging to the cluster
from foreground objects using the spectral signatures of massive
stars. That is, massive stars are hot, their spectra displaying
features attributable to transitions of hydrogen and helium; while
foreground low-mass stars are cool and have features attributable to
molecular CO.

In Table \ref{tab:lores} we list the stars in close proximity to the
RSGs $A$ and $B$ that were observed at low spectral resolution. We
also tabulate the photometry of these stars -- where none was
available in the literature or from our NIRC2 data, we obtained
photometry from the UKIDSS Galactic Plane Survey \citet{Lucas08}. We
assigned coarse spectral types to the objects based on the presence of
CO bandhead absorption -- those with CO were classified as late-type
(K/M); those without as early-type (O/B).

\begin{figure}
  \centering
  \includegraphics[width=8.5cm]{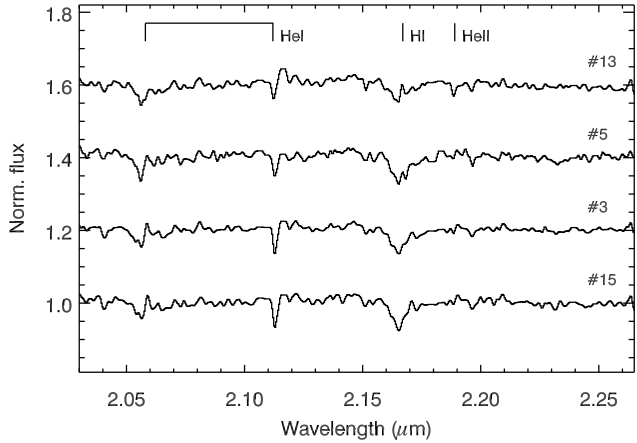}
  \caption{Low-resolution spectra of early-type stars in the
    cluster.\\ }
  \label{fig:bstars}
\end{figure}

In Fig.\ \ref{fig:bstars} we plot the spectra of the `early-type'
stars. All stars exhibit clear spectral features of Br-$\gamma$+HeI
(2.161,2.164), \hei-2.112\microns\ and \hei-2.058\microns. There is a
suggestion of \heii-2.189\microns\ absorption in the spectrum of \#13,
though the strength of this feature is comparable to the noise. To
spectral-type the stars, we compare to the spectral atlases of
\citet{Hanson96} and \citet{Hanson05}. The absence of any strong
\heii-2.189\microns, as well as a lack of \civ\ and \niii, suggests
that the stars cannot be any earlier than $\sim$O7. Also, the
comparable strengths of the Br-$\gamma$+\hei
(2.161,2.164)\microns\ suggests that the stars are no later than
$\sim$B3. At this spectral resolution, it is not possible to
distinguish between dwarfs and supergiants using the width of
Br-$\gamma$ as it blends with the adjacent \hei\ feature.

\begin{table}
  \centering
  \caption{Stars observed at low spectral resolution, with
    photometry and rough spectral types. Stellar identifications are
    taken from \citet{Vrba00}. Photometry is taken from UKIDSS, with
    the exception of $V-C$ \citep[from][]{Vrba96} and $V-5a$ (from
    NIRC2 photometry presented in this paper).\\ }
  \begin{tabular}{lcccc}
\hline
  Star  & J     & H     & K     & Spec type \\
\hline \hline
  V-C   & 12.47 & 11.30 & 10.45 &  K/M  \\
  V-1a  & 13.07 & 11.91 & 11.17 &  K/M  \\
  V-2   & 14.22 & 13.02 & 12.26 &  K/M  \\
  V-3   & 14.51 & 12.86 & 12.05 &  O/B  \\
  V-4a  & 14.63 & 13.25 & 12.38 &  K/M  \\
  V-5a  & --    & 15.19 & 14.41 &  O/B  \\
  V-13  & 14.25 & 12.68 & 11.98 &  O/B  \\
  V-15  & 13.67 & 12.12 & 11.36 &  O/B  \\
  \hline \\
  \end{tabular}
  \label{tab:lores}
\end{table}

\begin{figure}
  \centering
  \includegraphics[width=8.5cm]{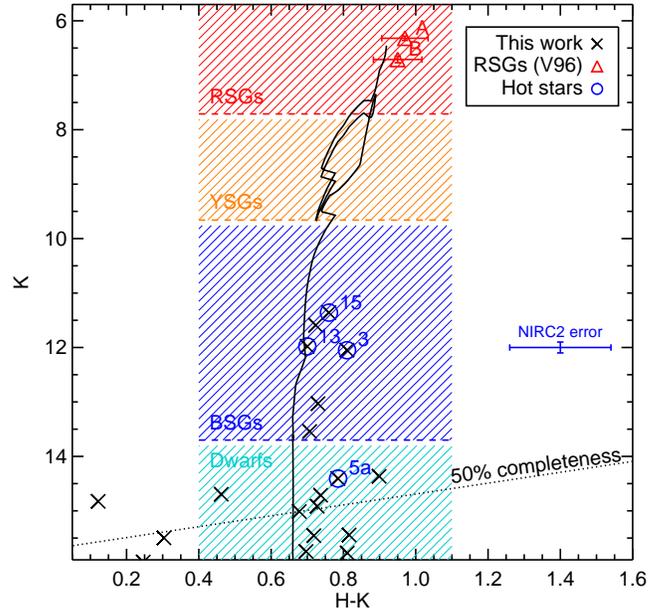}
  \caption{Colour-magnitude diagram for stars in the NIRC2
    field-of-view, including the photometry from V96. Overplotted is
    an 15Myr, solar-metallicity isochrone from \citet{Schaller92},
    using the distance and extinction derived in this paper. Indicated
    on the plot are the magitude ranges for RSGs, YSGs, BSGs and
    dwarfs in a cluster of that age, distance and extinction.\\ }
  \label{fig:colmag}
\end{figure}

In Fig.\ \ref{fig:colmag} we show the colour-magnitude diagram (CMD)
for the NIRC2 field. We also show the photometry of the three bright
stars from V96. A 15Myr isochrone from the non-rotating Geneva tracks
\citep{Schaller92}\footnote{We use the rotating models of
  \citet{Mey-Mae00} for most of our analysis, which at the moment are
  not available with instrumental colours.} is overplotted, using the
extinction and distance derived in Sect.\ \ref{sec:rsgs}.

In the panel we have indicated the apparent K-band magnitude ranges
for RSGs, Blue Supergiants (BSGs), Yellow Supergiants (YSGs) and dwarf
stars for a cluster with this age, distance and extinction, using the
\citet{Schaller92} models. They are defined as follows:

\begin{itemize}
\item RSGs: $T_{\rm eff} < 4,000K, L_{\star} > 10^4$\lsun
\item YSGs: $10,000K > T_{\rm eff} > 4,000K, L_{\star} > 10^4$\lsun
\item BSGs: $T_{\rm eff} > 10,000K, \log g < 3.5, L_{\star} >
  10^4$\lsun
\item Dwarfs\footnote{We use $\log g$ as a classification criterion in
  order to distinguish between main-sequence and post main-sequence
  objects. }: $\log g > 3.5, L_{\star} < 10^4$\lsun
\end{itemize}

The plot shows that, though there are two RSGs in the cluster, there
are no stars with brightesses consistent with being YSGs. There are
some objects with luminosities one would expect to see of BSGs --
however, several of these stars have been spectroscopically identified
as cool stars, and are therefore likely foreground objects. Stars
which show signatures of early spectral types are indicated in
Fig.\ \ref{fig:colmag} as circles. Overall we find three stars with
$K$-band magnitudes consistent with BSGs and one star consistent with
a main-sequence object.

\begin{figure}
  \centering
  \includegraphics[width=8.5cm]{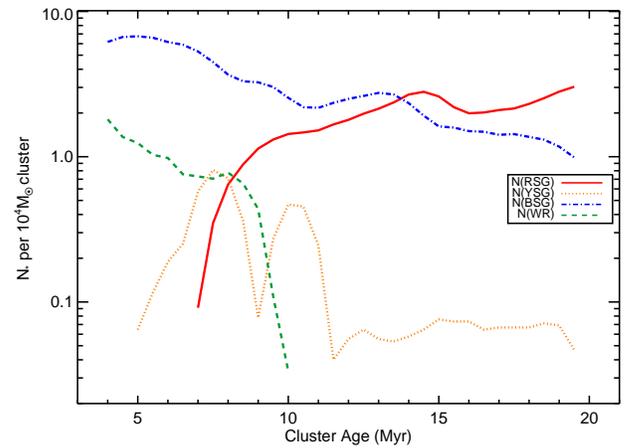}
  \caption{The relative numbers of BSGs, RSGs, YSGs and WRs in a
    coeval cluster of mass $10^{4}$\msun\ as a function of cluster
    age. \\ }
  \label{fig:starnum}
\end{figure}

In comparing the relative fractions of the different evolutionary
stages with those predicted by the evolutionary track, we find good
agreement. Using models with ages 13-17Myr from \citet{Schaller92} and
an Initial Mass Function (IMF) defined by \citet{Kroupa01}, RSGs are
predicted to outnumber YSGs by a factor of $\ga$10. The ratio of the
number of RSGs to BSGs is subject to large variations depending on the
age of the cluster, and can be between 0.1--10 in the age range we
determine for the cluster. This is also illustrated in
Fig.\ \ref{fig:starnum}, where we use the updated evolutionary models
of \citet{Mey-Mae00} which include stellar rotation to determine the
relative fractions of evolved massive stars in a coeval cluster as a
function of age.

\begin{table*}
  \begin{center}
  \scriptsize
  \caption{Post-supernova objects for which progenitor masses can be
    estimated from their host clusters. References: 1: \cite{Bibby08};
    2: \cite{Kouveliotou98}; 3: \cite{Muno06}; 4: \cite{Helfand07}; 5:
    \cite{Messineo08}; 6: \cite{G-H09}; 7: \cite{RSGC1paper};
    8: \cite{G-H08}; 9: \cite{Kouveliotou99}.\\   }
  \begin{tabular}{lcccc}
    \hline \hline
    Object [+ cluster] & $M_{\rm prog}$/\msun & Remnant & B
    ($\times$10$^{14}$G) & Ref. \\
    \hline
    SGR~1806-20        & 48$^{+20}_{-8}$      & Magnetar & 2-8 &
    1,2  \\  
    CXO~J164710.2-455216 [Wd~1] & 40$\pm$5   & Magnetar & $<$1.5 &
    3 \\ 
    IGR J18135-1751 [Cl~1813-18] & 20-30     & Pulsar Wind Nebula  &
    $\sim$1\footnote{This value is highly uncertain due to the lack of a
      measured period and spin-down rate, and may be an order of
      magnitude lower.}  & 4,5,6 \\
    AX~J1838-0655 [RSGC1] & 18$\pm$2         & Pulsar Wind Nebula  & 0.02   &
    7,8 \\ 
    SGR~1900+14        & 17$\pm$1       & Magnetar & 2-8 & This
    work, 9 \\
    \hline
    \label{tab:cl}
  \end{tabular}
  \end{center}
\end{table*}

Thus, we can say that we are unlikely to find any YSGs in a cluster
that contained only two RSGs, while the number of BSGs should be
within an order of magnitude of the number of RSGs. In addition, we
find that there should be approximately 10 times as many main-sequence
stars brighter than our 50\% completeness limit at $K=15$. While we
have only identified one such star here, there are many more B-dwarf
candidates in the NIRC2 field-of-view which occupy the relevant region
of the CMD and which await spectroscopic classification.

From this analysis, we conclude from the luminosities of the RSGs, the
absence of WRs, and the relative numbers of $N$(RSG)/$N$(BSG), that
Cl~1900+14 is fully consistent with being a 14$\pm$1\,Myr, coeval
cluster. By extrapolating over the rest of a Kroupa IMF down to
0.01\msun, we estimate that the total cluster mass of this cluster is
$\sim 10^{3}$\msun to within an order of magnitude.



\section{The initial mass of the magnetar's progenitor} \label{sec:disc}

To summarize the results of the previous section, in terms of the
cluster's age we find that the RSG luminosities are {\it uniquely} fit
by the rotating stellar evolutionary models at Solar metallicity
\cite{Mey-Mae00} for an age of 14$\pm$1Myr. Analysis of the fainter
stars indicate that the cluster is consistent with being a coeval
starburst to within the errors. No Wolf-Rayet stars are found, which
would imply star formation within the last $\sim$8Myr, while we find
relative numbers of hot/cool stars that are consistent with the model
predictions for a coeval 14Myr cluster. As the age of the cluster is
much greater than the lifetime of the magnetar ($\la10^4$yrs), we can
now estimate the mass of the magnetar's progenitor by determining the
mass of the most massive star that could still exist in a cluster of
this age. Using the same stellar evolution models as above we find
that the initial mass of the magnetar's progenitor was $M_{\rm prog} =
17\pm$1\msun\ (see Fig.\ \ref{fig:minmax}).

\begin{figure*}
  \includegraphics[width=8.9cm,bb=20 0 635 528]{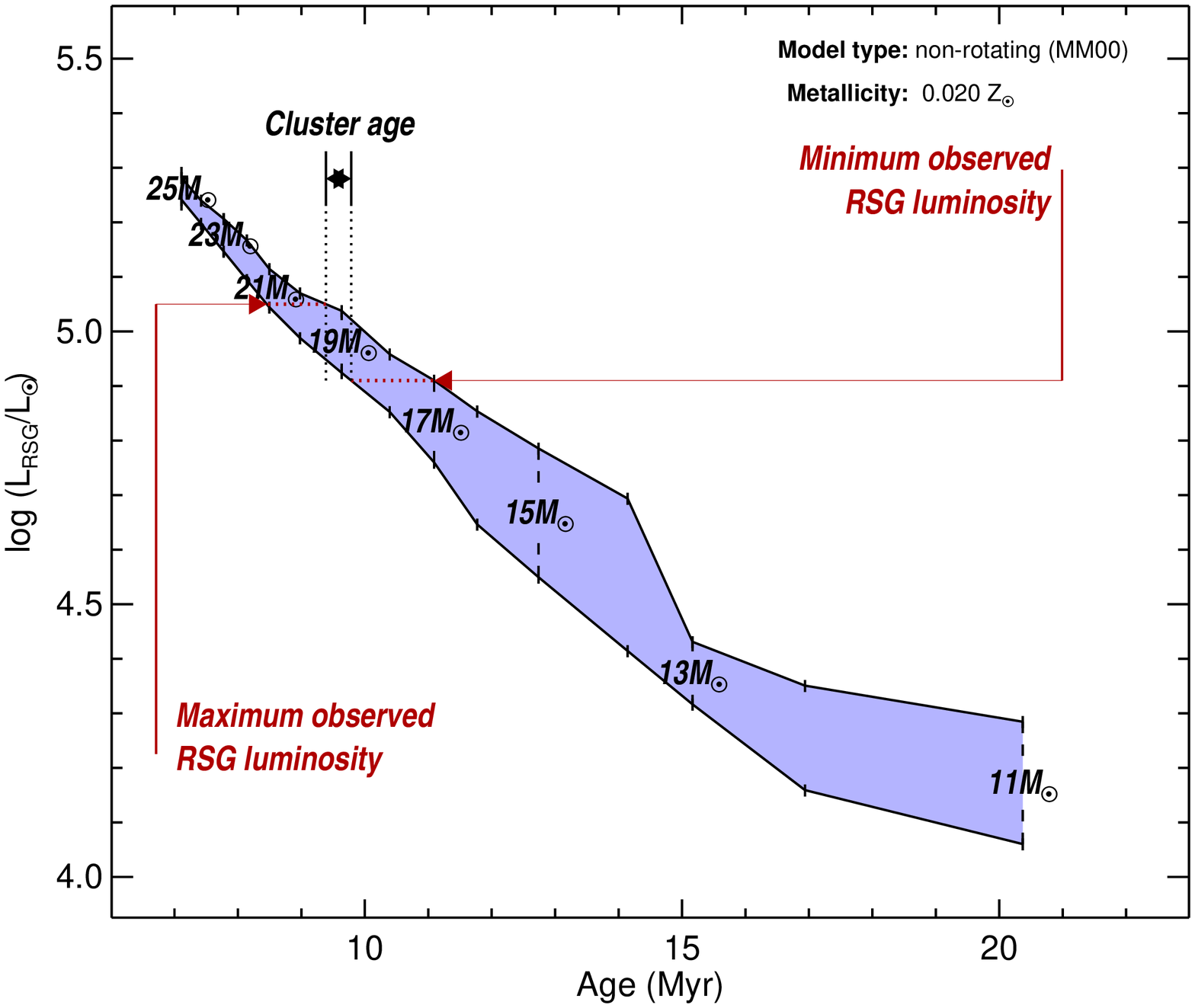}
  \includegraphics[width=8.9cm,bb=20 0 635 528]{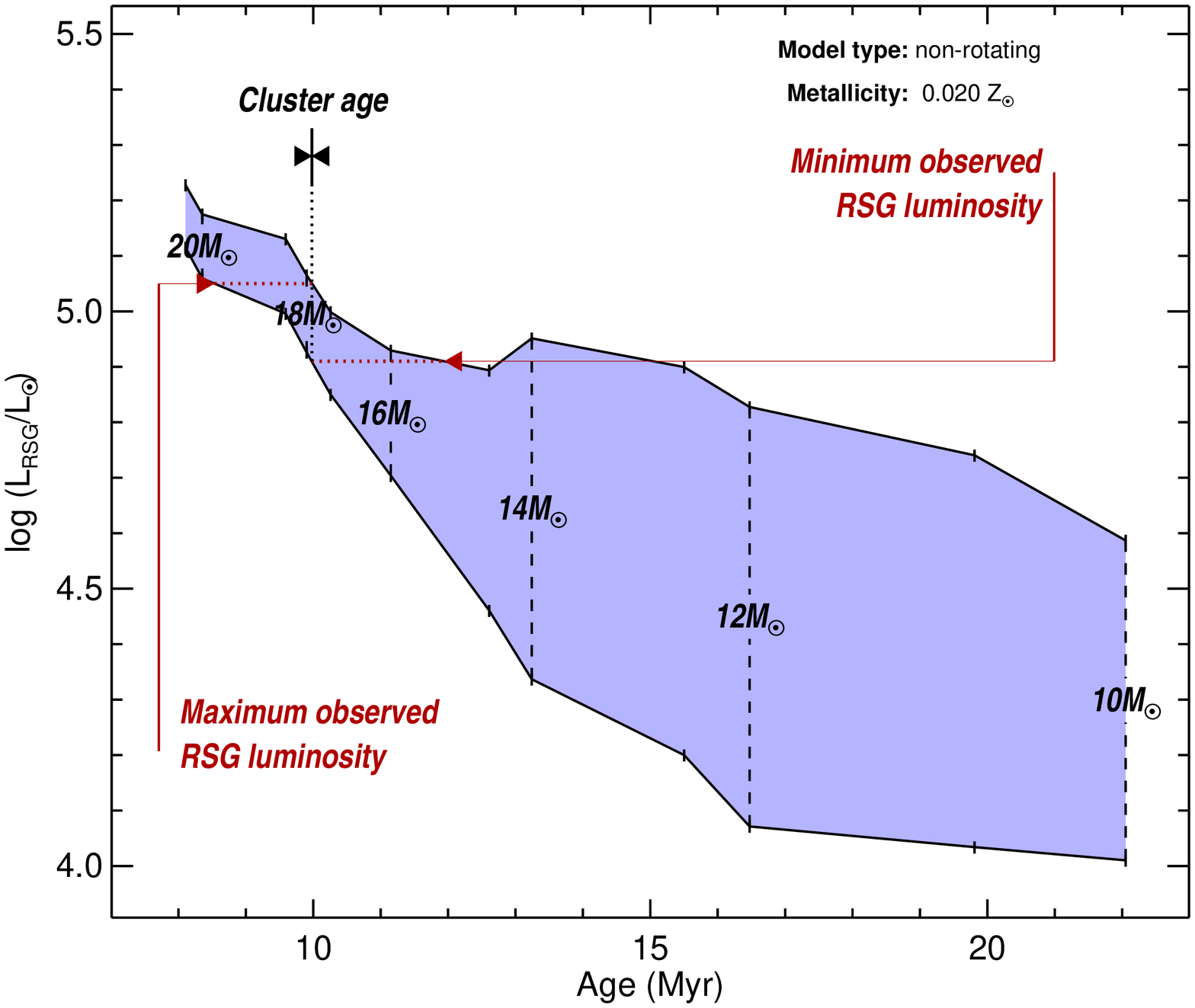}
  \includegraphics[width=8.9cm,bb=20 0 635 528]{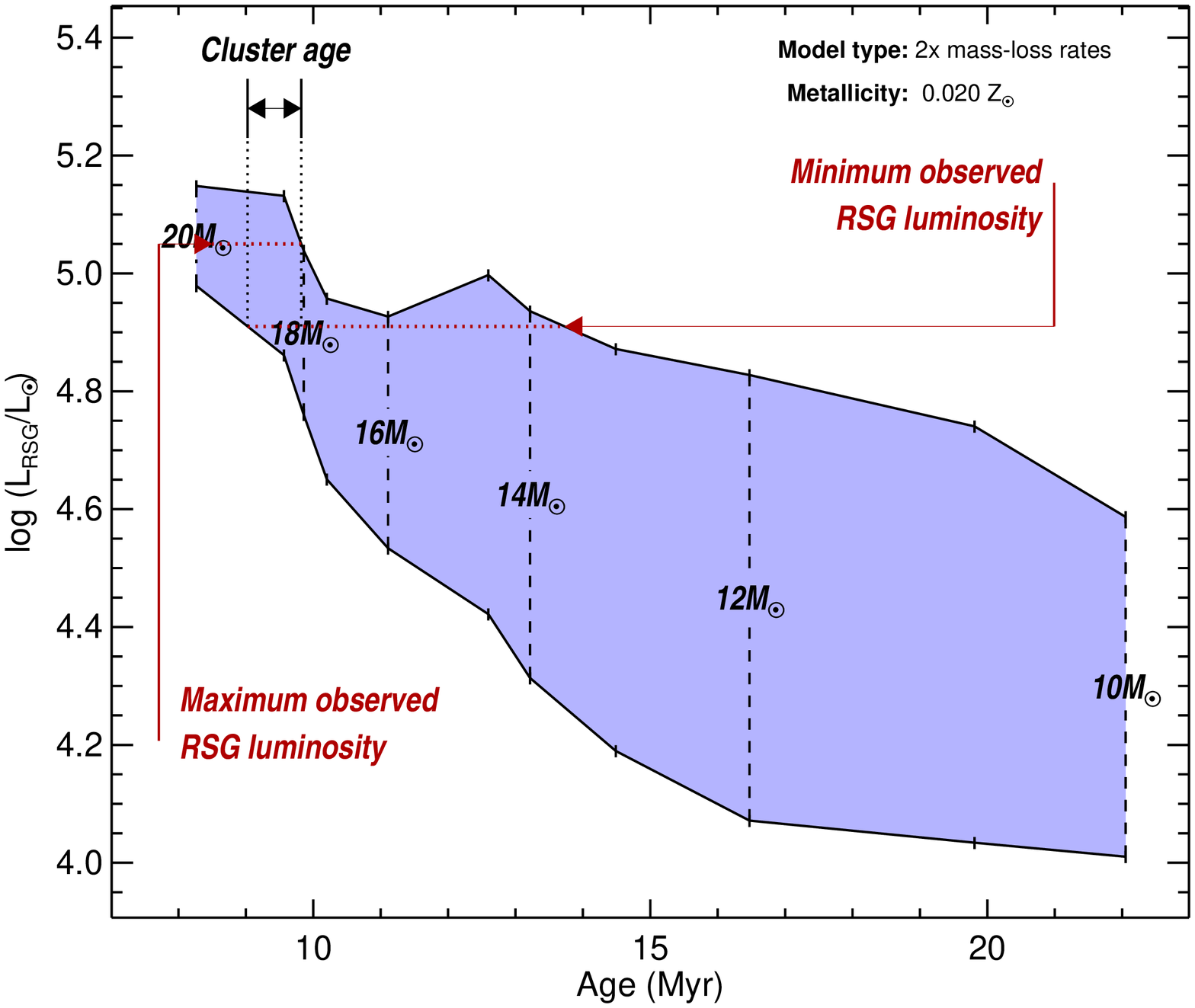}
  \includegraphics[width=8.9cm,bb=20 0 635 528]{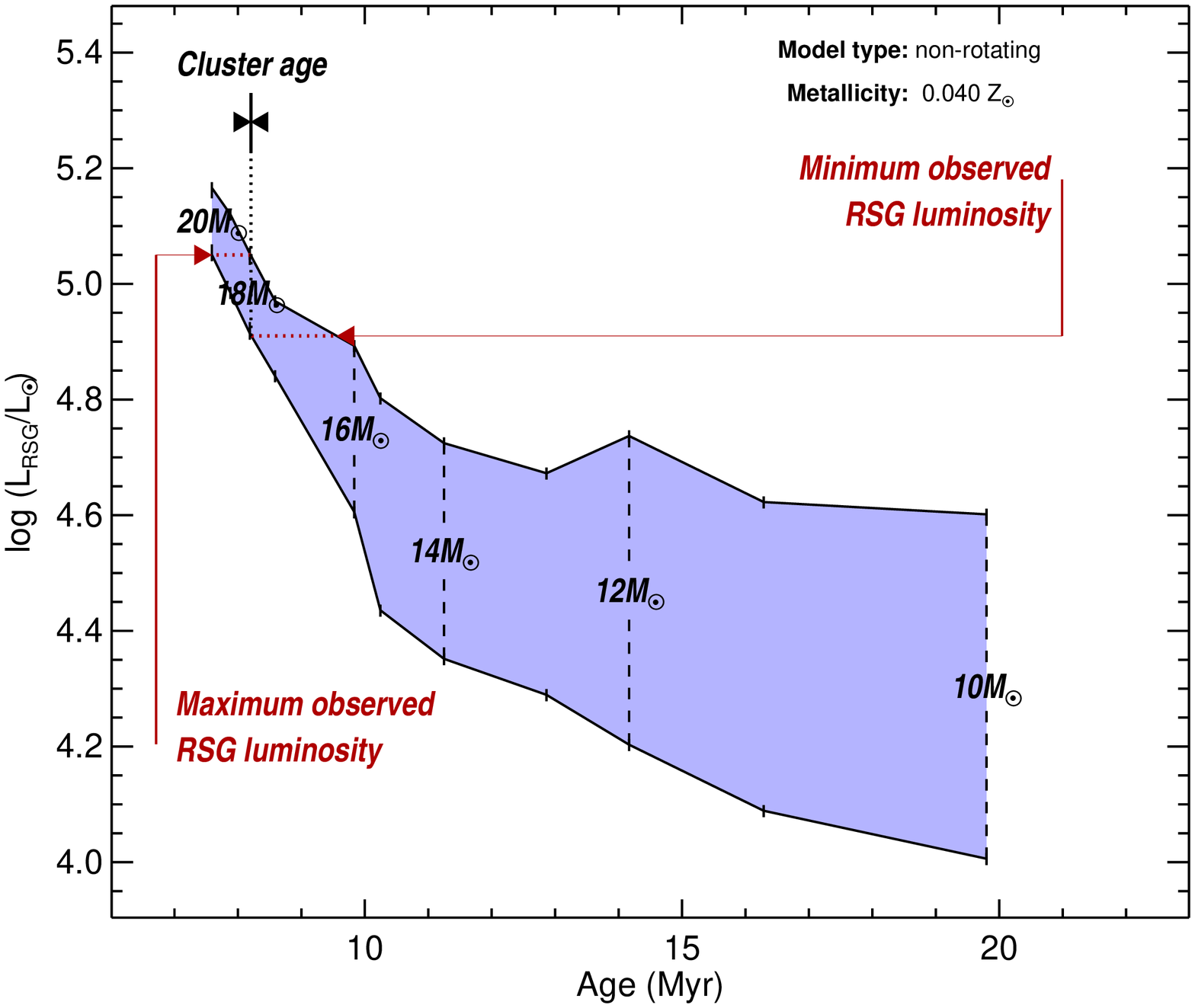}
  \includegraphics[width=8.9cm,bb=20 0 635 528]{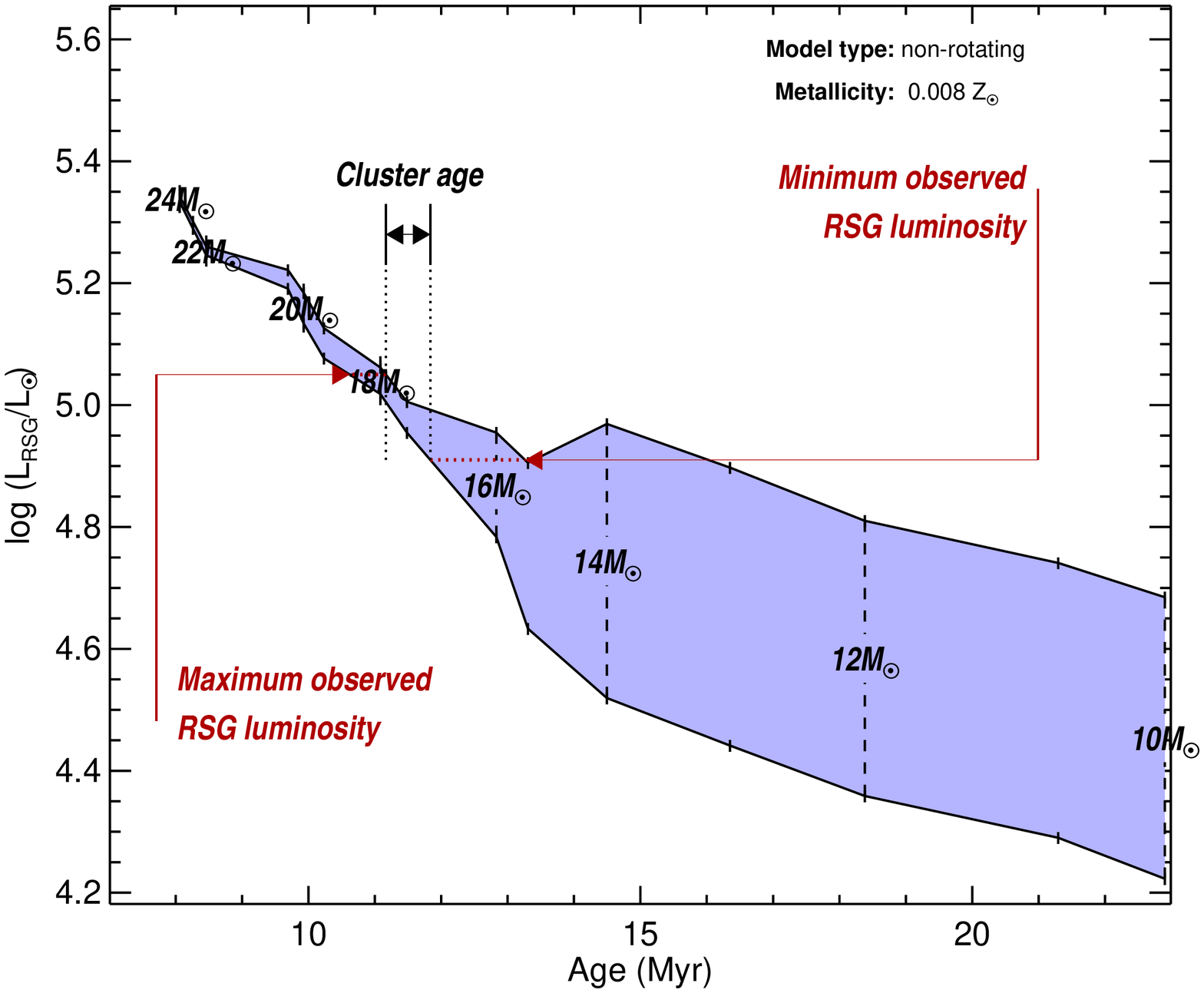}
  \caption{Similar figures to that of Fig.\ \ref{fig:minmax}. The
    figures illustrate the implied age of a cluster containing RSGs
    with the luminosity range we observe for Cl~1900+14 when different
    evolutionary models are used. Top left: non-rotating models
    \citep{Mey-Mae00}; top-right: obsolete non-rotating models
    \citep{Schaller92}; middle-left: same as top-right but with
    doubled mass-loss rates \citep{Meynet94}; middle-right: same as
    top-right but twice-Solar metallicity \citep{Schaerer93};
    bottom-left: same as top-right but for LMC metallicity
    \citep{Schaerer93}.  }
  \label{fig:diffmodels}
\end{figure*}

In Fig.\ \ref{fig:diffmodels} we examine the robustness of this result
by exploring the effects of using different evolutionary models to
determine the cluster age. In the top-left panel of
Fig.\ \ref{fig:diffmodels} we show the result of using the updated
Geneva models which do not include rotation \citep{Mey-Mae00}. As
discussed by Meynet \& Maeder, one effect of including rotation in
their evolutionary code is that the post main-sequence massive stars
become more luminous. Hence, for a given stellar luminosity, the
inferred age from rotating models is larger than for non-rotating
models. We see from the top-left panel of Fig.\ \ref{fig:diffmodels}
that when the non-rotating models are used the cluster age is reduced
to 9-10~Myr. {\it However}, the implied initial masses of the RSGs is
altered only slightly -- $\sim$19\msun, as opposed to 16-17\msun\ from
the rotating models.

In the remaining panels of Fig.\ \ref{fig:diffmodels} we explore the
use of earlier models of the Geneva group, which do not include
rotation but which do include convective overshooting. We see that,
while inferred cluster ages are systematically lower than those
determined using rotating models, the RSG masses are consistently in
the region of 17-19\msun. This is the case whether the metallicity is
Solar, twice-Solar or Solar/2.5. Similarly, doubling the stellar
mass-loss rates has little impact. 

From this analysis we conclude that the version of evolutionary model
used to calculate the initial mass of the magnetar's progenitor has
very little impact on the value derived. From the contemporary models
which include rotation we find $M_{\rm prog} = 17 \pm 1$\msun; while
from earlier variations of the Geneva evolutionary code we
consistently find masses in the range of 17-19\msun. Therefore, we
find that any systematic uncertainty in $M_{\rm prog}$ must be small,
of order 1-2\msun. To account for systematic uncertainties we adopt an
error of $\pm$2\msun\ on the derived progenitor mass.

\subsection{The possible role of binarity}

We now consider the role that binarity may play in magnetar
production. To recap, one supporting piece of evidence cited by
\citet{Gaensler05} for high progenitor masses of magnetars was from
the numerical simulations of \citet{Heger05}. These authors found that
the presence of internal magnetic fields may work to transport angular
momentum away from the core, particularly during the transition to
core He burning, as the star adjusts to the large differential
rotation between the core and the inflated envelope of the RSG
phase. A progenitor with initial mass similar to that we derive for
SGR~1900+14 would almost certainly pass through the RSG phase as a
{\it single} star. However, if a star were to evolve in a close binary
system, it has been shown that the effects of mass-transfer onto a
companion star significantly reduce the RSG lifetime
\citep{Eldridge08}. This enables the core to retain angular momentum
through to SN, enough perhaps to greatly amplify the resulting neutron
star's magnetic field through the dynamo mechanism in the first few
seconds of its life. In addition, the presence of a secondary star may
serve to add angular momentum to the primary. We note however that in
this scenario the secondary star would remain after the primary had
gone SN. As yet, there is no compelling evidence for an
optical/near-IR counterpart to SGR1900+14 \citep[a faint candidate
  with $K=19.21$ was suggested by][]{Testa08}.

\subsubsection{Was the progenitor the product of a merger?}
If the magnetar's progenitor star was part of a binary system that
merged before the primary went SN, this could explain the lack of any
obvious optical/IR counterpart. Indeed, such a process may seem
appealing from the point-of-view of magnetar production; it would
increase the terminal mass of the star, while it would also likely
spin-up the core. If such a merger did occur it must have been at a
time shortly before the SN explosion that created the
magnetar. Otherwise, the lifetime of the resulting more-massive star
would be significantly shortened and would have gone SN many millions
of years ago. Hence, the time since SN would be much greater than
current estimates of magnetar lifetimes ($10^{4}$yrs).

Such an explanation may seem contrived, however a similar scenario has
been proposed for the progenitor of SN1987A, where two stars of masses
$\sim$15\msun\ and $\sim$5\msun\ merged around 20,000yrs prior to SN
\citep[see e.g.][]{Podsiadlowski92}. It is commonly thought that
mergers of stars in binary systems are far more likely to occur in
systems with large mass ratios. In order to merge, the primary must
evolve off the main-sequence such that the secondary is enveloped in
the primary's inflated atmosphere, whereupon the two stars spiral
downwards toward the centre of mass. Hence, if a merger did occur in
the progenitor of SGR1900+14, it is unlikely that the pre-merger
system consisted of two 17\msun\ stars and was more likely a
17\msun\ primary and a secondary of mass $\la$10\msun. 

We cannot discount the possibility that the progenitor of SGR1900+14
experienced a merger, nor can we rule out that such mergers are a
necessary ingredient in the magnetar production mechanism. However,
the evidence for such a merger is limited at best. While the magnetar
is at the centre of a large-scale ring nebula \citep{Wachter08}, and a
similar ring around SN1987A is proposed to be the product of a merger
prior to SN \citep{M-P06}, mergers are not the only mechanism to
produce such rings. This alone is not compelling evidence for a merger
event for the progenitor of SGR1900+14.

\subsection{Comparison with other neutron stars with known
  progenitor masses}

How does our mass measurement of 17\msun\ compare to other post-SN
objects with progenitor mass estimates? In Table \ref{tab:cl} we list
all known young clusters associated with neutron stars. As well as the
three clusters containing magnetars, we also list the two recent
discoveries of clusters associated with Pulsar Wind Nebulae (PWNe) --
Cl~$1813-13$ \citep{Helfand07,Messineo08,G-H09}, and RSGC1
\citep{RSGC1paper,G-H08}. Prior to our current result, it could be
argued from these data that there is a connection between magnetic
field strength $B$ and progenitor mass. However, the inclusion of
SGR~1900+14 -- the object with the lowest progenitor mass of the
sample, but whose magnetic field is as large as any other on the list
-- appears to end any notion of a relation between $B$ and $M_{\rm
  prog}$.  As such, our result provides a strong challenge to the
hypothesis that magnetars decend from very massive stars --
specifically, those stars massive enough to avoid the RSG phase and
the associated core spin-down \citep{Heger05}\footnote{However, see
  also \citet{Ott06} for calculations on neutron star birth
  periods.}. From our current understanding of stellar physics, it is
not possible for a 17\msun\ single star to shed enough of its
hydrogen-rich envelope on the main-sequence to avoid the RSG phase
\citep{Mey-Mae00} (The influence of binarity on the evolution of a
star of similar mass was explored in the previous section).

If magnetars can be produced from stars which will inevitably suffer
core spin-down during their evolution, then perhaps stellar rotation,
and in turn initial stellar mass, are not the primary factors in the
production of extreme magnetic fields in neutron stars. An alternative
theory to the dynamo mechanism is the `fossil-field' scenario, whereby
a seed $B$-field is inherited by a newly-born star from its natal
molecular cloud \citep[e.g.][]{Ferrario05}. This explanation is
preferred from studies of the energetics of SN remnants associated
with magnetars, in which no evidence has been found for the extra
energy boost provided by the neutron star's rapid spin-down (such as
predicted by the dynamo scenario) \citep{V-K06}. However, while a
handful of massive stars have recently been observed to have magnetic
field strengths of $\sim 10^{3}$G \citep{Donati06,Bouret08}, it has
been noted that flux conservation alone is not sufficient to amplify
the B-field to that of magnetar levels \citep{Spruit08}, and no
current theory exists for how such fields evolve with the star up to
the point of supernova.

\section{Summary \& conclusions}
We present an imaging and spectroscopic study of the host cluster of
the magnetar SGR1900+14, with the purpose of deriving the initial mass
of the magnetar's progenitor. From analysis of the two bright Red
Supergiants in the cluster we determine a kinematic distance of
12.5$\pm$1.7kpc and extinction of $A_{\rm V} =12.9\pm0.5$, which is in
good agreement with that previously derived for the magnetar. From the
luminosities of the RSGs we determine an age for the cluster of
14$\pm$1 Myr, and a study of the fainter stellar population of the
cluster reveals that it is consistent with being coeval to within the
errors. Assuming that the SN that created the magnetar occured very
recently in comparison to the age of the cluster, we derive an initial
mass of the magnetar's progenitor of $M_{\rm prog} = 17 \pm
2$\msun. We find this estimate is insensitive to parameters such as
metallicity and the type of evolutionary models used. This result is
significantly lower than for other magnetars with progenitor mass
estimates, and challenges the hypothesis that very high initial masses
($\ga40$\msun) are required to produce magnetars. Instead, we suggest
that some other initial parameter, such as magnetic field strength at
birth or the presence of a close binary companion, may be the dominant
factor in producing a super-strong magnetic field.

\acknowledgments Acknowledgments: we thank Jim Hinton and John
Eldridge for useful discussion, and the anonymous referee for helpful
comments and suggestions. This work makes use of the UKIDSS survey;
the UKIDSS project is defined in \citet{Lawrence07}. UKIDSS uses the
UKIRT Wide Field Camera \citep[WFCAM][]{Casali07} and a photometric
system described in \citet{Hewitt06}. The pipeline processing and
science archive are described in and \citet{Hambly08}. The material in
this work is supported by NASA under award NNG~05-GC37G, through the
Long-Term Space Astrophysics program. This research was performed in
the Rochester Imaging Detector Laboratory with support from a NYSTAR
Faculty Development Program grant. Part of the data presented here
were obtained at the W.\ M.\ Keck Observatory, which is operated as a
scientific partnership among the California Institute of Technology,
the University of California, and the National Aeronautics and Space
Administration. The Observatory was made possible by the generous
financial support of the W.\ M.\ Keck Foundation. This research has
made use of the IDL software package and the GSFC IDL library.

\bibliographystyle{/fat/Data/bibtex/apj}
\bibliography{/fat/Data/bibtex/biblio}

\begin{thebibliography}{50}
\expandafter\ifx\csname natexlab\endcsname\relax\def\natexlab#1{#1}\fi

\bibitem[{{Bibby} {et~al.}(2008){Bibby}, {Crowther}, {Furness}, \&
  {Clark}}]{Bibby08}
{Bibby}, J.~L., {Crowther}, P.~A., {Furness}, J.~P., \& {Clark}, J.~S. 2008,
  \mnras, 386, L23

\bibitem[{{Bouret} {et~al.}(2008){Bouret}, {Donati}, {Martins}, {Escolano},
  {Marcolino}, {Lanz}, \& {Howarth}}]{Bouret08}
{Bouret}, J.-C., {Donati}, J.-F., {Martins}, F., {Escolano}, C., {Marcolino},
  W., {Lanz}, T., \& {Howarth}, I.~D. 2008, \mnras, 389, 75

\bibitem[{{Brand} \& {Blitz}(1993)}]{B-B93}
{Brand}, J. \& {Blitz}, L. 1993, \aap, 275, 67

\bibitem[{{Casali} {et~al.}(2007){Casali}, {Adamson}, {Alves de Oliveira},
  {Almaini}, {Burch}, {Chuter}, {Elliot}, {Folger}, {Foucaud}, {Hambly},
  {Hastie}, {Henry}, {Hirst}, {Irwin}, {Ives}, {Lawrence}, {Laidlaw}, {Lee},
  {Lewis}, {Lunney}, {McLay}, {Montgomery}, {Pickup}, {Read}, {Rees}, {Robson},
  {Sekiguchi}, {Vick}, {Warren}, \& {Woodward}}]{Casali07}
{Casali}, M., {Adamson}, A., {Alves de Oliveira}, C., {Almaini}, O., {Burch},
  K., {Chuter}, T., {Elliot}, J., {Folger}, M., {Foucaud}, S., {Hambly}, N.,
  {Hastie}, M., {Henry}, D., {Hirst}, P., {Irwin}, M., {Ives}, D., {Lawrence},
  A., {Laidlaw}, K., {Lee}, D., {Lewis}, J., {Lunney}, D., {McLay}, S.,
  {Montgomery}, D., {Pickup}, A., {Read}, M., {Rees}, N., {Robson}, I.,
  {Sekiguchi}, K., {Vick}, A., {Warren}, S., \& {Woodward}, B. 2007, \aap, 467,
  777

\bibitem[{{Davies} {et~al.}(2007){Davies}, {Figer}, {Kudritzki}, {MacKenty},
  {Najarro}, \& {Herrero}}]{RSGC2paper}
{Davies}, B., {Figer}, D.~F., {Kudritzki}, R.-P., {MacKenty}, J., {Najarro},
  F., \& {Herrero}, A. 2007, \apj, 671, 781

\bibitem[{{Davies} {et~al.}(2008){Davies}, {Figer}, {Law}, {Kudritzki},
  {Najarro}, {Herrero}, \& {MacKenty}}]{RSGC1paper}
{Davies}, B., {Figer}, D.~F., {Law}, C.~J., {Kudritzki}, R.-P., {Najarro}, F.,
  {Herrero}, A., \& {MacKenty}, J.~W. 2008, \apj, 676, 1016

\bibitem[{{Donati} {et~al.}(2006){Donati}, {Howarth}, {Bouret}, {Petit},
  {Catala}, \& {Landstreet}}]{Donati06}
{Donati}, J.-F., {Howarth}, I.~D., {Bouret}, J.-C., {Petit}, P., {Catala}, C.,
  \& {Landstreet}, J. 2006, \mnras, 365, L6

\bibitem[{{Duncan} \& {Thompson}(1992)}]{D-T92}
{Duncan}, R.~C. \& {Thompson}, C. 1992, \apjl, 392, L9

\bibitem[{{Eldridge} {et~al.}(2008){Eldridge}, {Izzard}, \&
  {Tout}}]{Eldridge08}
{Eldridge}, J.~J., {Izzard}, R.~G., \& {Tout}, C.~A. 2008, \mnras, 384, 1109

\bibitem[{{Ferrario} \& {Wickramasinghe}(2005)}]{Ferrario05}
{Ferrario}, L. \& {Wickramasinghe}, D.~T. 2005, \mnras, 356, 615

\bibitem[{{Figer} {et~al.}(2003){Figer}, {Gilmore}, {Kim}, {Morris}, {Becklin},
  {McLean}, {Gilbert}, {Graham}, {Larkin}, {Levenson}, \& {Teplitz}}]{Figer03}
{Figer}, D.~F., {Gilmore}, D., {Kim}, S.~S., {Morris}, M., {Becklin}, E.~E.,
  {McLean}, I.~S., {Gilbert}, A.~M., {Graham}, J.~R., {Larkin}, J.~E.,
  {Levenson}, N.~A., \& {Teplitz}, H.~I. 2003, \apj, 599, 1139

\bibitem[{{Figer} {et~al.}(2005){Figer}, {Najarro}, {Geballe}, {Blum}, \&
  {Kudritzki}}]{Figer05sgr}
{Figer}, D.~F., {Najarro}, F., {Geballe}, T.~R., {Blum}, R.~D., \& {Kudritzki},
  R.~P. 2005, \apjl, 622, L49

\bibitem[{{Gaensler} {et~al.}(2005{\natexlab{a}}){Gaensler}, {Kouveliotou},
  {Gelfand}, {Taylor}, {Eichler}, {Wijers}, {Granot}, {Ramirez-Ruiz},
  {Lyubarsky}, {Hunstead}, {Campbell-Wilson}, {van der Horst}, {McLaughlin},
  {Fender}, {Garrett}, {Newton-McGee}, {Palmer}, {Gehrels}, \&
  {Woods}}]{Gaensler05}
{Gaensler}, B.~M., {Kouveliotou}, C., {Gelfand}, J.~D., {Taylor}, G.~B.,
  {Eichler}, D., {Wijers}, R.~A.~M.~J., {Granot}, J., {Ramirez-Ruiz}, E.,
  {Lyubarsky}, Y.~E., {Hunstead}, R.~W., {Campbell-Wilson}, D., {van der
  Horst}, A.~J., {McLaughlin}, M.~A., {Fender}, R.~P., {Garrett}, M.~A.,
  {Newton-McGee}, K.~J., {Palmer}, D.~M., {Gehrels}, N., \& {Woods}, P.~M.
  2005{\natexlab{a}}, \nat, 434, 1104

\bibitem[{{Gaensler} {et~al.}(2005{\natexlab{b}}){Gaensler},
  {McClure-Griffiths}, {Oey}, {Haverkorn}, {Dickey}, \&
  {Green}}]{Gaensler05apj}
{Gaensler}, B.~M., {McClure-Griffiths}, N.~M., {Oey}, M.~S., {Haverkorn}, M.,
  {Dickey}, J.~M., \& {Green}, A.~J. 2005{\natexlab{b}}, \apjl, 620, L95

\bibitem[{{Gotthelf} \& {Halpern}(2008)}]{G-H08}
{Gotthelf}, E.~V. \& {Halpern}, J.~P. 2008, \apj, 681, 515

\bibitem[{{Gotthelf} \& {Halpern}(2009)}]{G-H09}
---. 2009, \apjl, 700, L158

\bibitem[{{Hambly} {et~al.}(2008){Hambly}, {Collins}, {Cross}, {Mann}, {Read},
  {Sutorius}, {Bond}, {Bryant}, {Emerson}, {Lawrence}, {Rimoldini}, {Stewart},
  {Williams}, {Adamson}, {Hirst}, {Dye}, \& {Warren}}]{Hambly08}
{Hambly}, N.~C., {Collins}, R.~S., {Cross}, N.~J.~G., {Mann}, R.~G., {Read},
  M.~A., {Sutorius}, E.~T.~W., {Bond}, I., {Bryant}, J., {Emerson}, J.~P.,
  {Lawrence}, A., {Rimoldini}, L., {Stewart}, J.~M., {Williams}, P.~M.,
  {Adamson}, A., {Hirst}, P., {Dye}, S., \& {Warren}, S.~J. 2008, \mnras, 384,
  637

\bibitem[{{Hanson} {et~al.}(1996){Hanson}, {Conti}, \& {Rieke}}]{Hanson96}
{Hanson}, M.~M., {Conti}, P.~S., \& {Rieke}, M.~J. 1996, \apjs, 107, 281

\bibitem[{{Hanson} {et~al.}(2005){Hanson}, {Kudritzki}, {Kenworthy}, {Puls}, \&
  {Tokunaga}}]{Hanson05}
{Hanson}, M.~M., {Kudritzki}, R.-P., {Kenworthy}, M.~A., {Puls}, J., \&
  {Tokunaga}, A.~T. 2005, \apjs, 161, 154

\bibitem[{{Heger} {et~al.}(2005){Heger}, {Woosley}, \& {Spruit}}]{Heger05}
{Heger}, A., {Woosley}, S.~E., \& {Spruit}, H.~C. 2005, \apj, 626, 350

\bibitem[{{Helfand} {et~al.}(2007){Helfand}, {Gotthelf}, {Halpern}, {Camilo},
  {Semler}, {Becker}, \& {White}}]{Helfand07}
{Helfand}, D.~J., {Gotthelf}, E.~V., {Halpern}, J.~P., {Camilo}, F., {Semler},
  D.~R., {Becker}, R.~H., \& {White}, R.~L. 2007, \apj, 665, 1297

\bibitem[{{Hewett} {et~al.}(2006){Hewett}, {Warren}, {Leggett}, \&
  {Hodgkin}}]{Hewitt06}
{Hewett}, P.~C., {Warren}, S.~J., {Leggett}, S.~K., \& {Hodgkin}, S.~T. 2006,
  \mnras, 367, 454

\bibitem[{{Kaplan} {et~al.}(2002){Kaplan}, {Kulkarni}, {Frail}, \& {van
  Kerkwijk}}]{Kaplan02}
{Kaplan}, D.~L., {Kulkarni}, S.~R., {Frail}, D.~A., \& {van Kerkwijk}, M.~H.
  2002, \apj, 566, 378

\bibitem[{{Kouveliotou} {et~al.}(1998){Kouveliotou}, {Dieters}, {Strohmayer},
  {van Paradijs}, {Fishman}, {Meegan}, {Hurley}, {Kommers}, {Smith}, {Frail},
  \& {Murakami}}]{Kouveliotou98}
{Kouveliotou}, C., {Dieters}, S., {Strohmayer}, T., {van Paradijs}, J.,
  {Fishman}, G.~J., {Meegan}, C.~A., {Hurley}, K., {Kommers}, J., {Smith}, I.,
  {Frail}, D., \& {Murakami}, T. 1998, \nat, 393, 235

\bibitem[{{Kouveliotou} {et~al.}(1994){Kouveliotou}, {Fishman}, {Meegan},
  {Paciesas}, {van Paradijs}, {Norris}, {Preece}, {Briggs}, {Horack},
  {Pendleton}, \& {Green}}]{Kouveliotou94}
{Kouveliotou}, C., {Fishman}, G.~J., {Meegan}, C.~A., {Paciesas}, W.~S., {van
  Paradijs}, J., {Norris}, J.~P., {Preece}, R.~D., {Briggs}, M.~S., {Horack},
  J.~M., {Pendleton}, G.~H., \& {Green}, D.~A. 1994, \nat, 368, 125

\bibitem[{{Kouveliotou} {et~al.}(1999){Kouveliotou}, {Strohmayer}, {Hurley},
  {van Paradijs}, {Finger}, {Dieters}, {Woods}, {Thompson}, \&
  {Duncan}}]{Kouveliotou99}
{Kouveliotou}, C., {Strohmayer}, T., {Hurley}, K., {van Paradijs}, J.,
  {Finger}, M.~H., {Dieters}, S., {Woods}, P., {Thompson}, C., \& {Duncan},
  R.~C. 1999, \apjl, 510, L115

\bibitem[{{Kroupa}(2001)}]{Kroupa01}
{Kroupa}, P. 2001, \mnras, 322, 231

\bibitem[{{Lawrence} {et~al.}(2007){Lawrence}, {Warren}, {Almaini}, {Edge},
  {Hambly}, {Jameson}, {Lucas}, {Casali}, {Adamson}, {Dye}, {Emerson},
  {Foucaud}, {Hewett}, {Hirst}, {Hodgkin}, {Irwin}, {Lodieu}, {McMahon},
  {Simpson}, {Smail}, {Mortlock}, \& {Folger}}]{Lawrence07}
{Lawrence}, A., {Warren}, S.~J., {Almaini}, O., {Edge}, A.~C., {Hambly}, N.~C.,
  {Jameson}, R.~F., {Lucas}, P., {Casali}, M., {Adamson}, A., {Dye}, S.,
  {Emerson}, J.~P., {Foucaud}, S., {Hewett}, P., {Hirst}, P., {Hodgkin}, S.~T.,
  {Irwin}, M.~J., {Lodieu}, N., {McMahon}, R.~G., {Simpson}, C., {Smail}, I.,
  {Mortlock}, D., \& {Folger}, M. 2007, \mnras, 379, 1599

\bibitem[{{Levesque} {et~al.}(2005){Levesque}, {Massey}, {Olsen}, {Plez},
  {Josselin}, {Maeder}, \& {Meynet}}]{Levesque05}
{Levesque}, E.~M., {Massey}, P., {Olsen}, K.~A.~G., {Plez}, B., {Josselin}, E.,
  {Maeder}, A., \& {Meynet}, G. 2005, \apj, 628, 973

\bibitem[{{Lucas} {et~al.}(2008){Lucas}, {Hoare}, {Longmore}, {Schr{\"o}der},
  {Davis}, {Adamson}, {Bandyopadhyay}, {de Grijs}, {Smith}, {Gosling},
  {Mitchison}, {G{\'a}sp{\'a}r}, {Coe}, {Tamura}, {Parker}, {Irwin}, {Hambly},
  {Bryant}, {Collins}, {Cross}, {Evans}, {Gonzalez-Solares}, {Hodgkin},
  {Lewis}, {Read}, {Riello}, {Sutorius}, {Lawrence}, {Drew}, {Dye}, \&
  {Thompson}}]{Lucas08}
{Lucas}, P.~W., {Hoare}, M.~G., {Longmore}, A., {Schr{\"o}der}, A.~C., {Davis},
  C.~J., {Adamson}, A., {Bandyopadhyay}, R.~M., {de Grijs}, R., {Smith}, M.,
  {Gosling}, A., {Mitchison}, S., {G{\'a}sp{\'a}r}, A., {Coe}, M., {Tamura},
  M., {Parker}, Q., {Irwin}, M., {Hambly}, N., {Bryant}, J., {Collins}, R.~S.,
  {Cross}, N., {Evans}, D.~W., {Gonzalez-Solares}, E., {Hodgkin}, S., {Lewis},
  J., {Read}, M., {Riello}, M., {Sutorius}, E.~T.~W., {Lawrence}, A., {Drew},
  J.~E., {Dye}, S., \& {Thompson}, M.~A. 2008, \mnras, 391, 136

\bibitem[{{Mereghetti}(2008)}]{Mereghetti08}
{Mereghetti}, S. 2008, \aapr, 15, 225

\bibitem[{{Messineo} {et~al.}(2008){Messineo}, {Figer}, {Davies}, {Rich},
  {Valenti}, \& {Kudritzki}}]{Messineo08}
{Messineo}, M., {Figer}, D.~F., {Davies}, B., {Rich}, R.~M., {Valenti}, E., \&
  {Kudritzki}, R.~P. 2008, \apjl, 683, L155

\bibitem[{{Meynet} \& {Maeder}(2000)}]{Mey-Mae00}
{Meynet}, G. \& {Maeder}, A. 2000, \aap, 361, 101

\bibitem[{{Meynet} {et~al.}(1994){Meynet}, {Maeder}, {Schaller}, {Schaerer}, \&
  {Charbonnel}}]{Meynet94}
{Meynet}, G., {Maeder}, A., {Schaller}, G., {Schaerer}, D., \& {Charbonnel}, C.
  1994, \aaps, 103, 97

\bibitem[{{Morris} \& {Podsiadlowski}(2006)}]{M-P06}
{Morris}, T. \& {Podsiadlowski}, P. 2006, \mnras, 365, 2

\bibitem[{{Muno} {et~al.}(2006){Muno}, {Clark}, {Crowther}, {Dougherty}, {de
  Grijs}, {Law}, {McMillan}, {Morris}, {Negueruela}, {Pooley}, {Portegies
  Zwart}, \& {Yusef-Zadeh}}]{Muno06}
{Muno}, M.~P., {Clark}, J.~S., {Crowther}, P.~A., {Dougherty}, S.~M., {de
  Grijs}, R., {Law}, C., {McMillan}, S.~L.~W., {Morris}, M.~R., {Negueruela},
  I., {Pooley}, D., {Portegies Zwart}, S., \& {Yusef-Zadeh}, F. 2006, \apjl,
  636, L41

\bibitem[{{Ott} {et~al.}(2006){Ott}, {Burrows}, {Thompson}, {Livne}, \&
  {Walder}}]{Ott06}
{Ott}, C.~D., {Burrows}, A., {Thompson}, T.~A., {Livne}, E., \& {Walder}, R.
  2006, \apjs, 164, 130

\bibitem[{{Podsiadlowski} {et~al.}(1992){Podsiadlowski}, {Joss}, \&
  {Hsu}}]{Podsiadlowski92}
{Podsiadlowski}, P., {Joss}, P.~C., \& {Hsu}, J.~J.~L. 1992, \apj, 391, 246

\bibitem[{{Rieke} \& {Lebofsky}(1985)}]{R-L85}
{Rieke}, G.~H. \& {Lebofsky}, M.~J. 1985, \apj, 288, 618

\bibitem[{{Schaerer} {et~al.}(1993){Schaerer}, {Charbonnel}, {Meynet},
  {Maeder}, \& {Schaller}}]{Schaerer93}
{Schaerer}, D., {Charbonnel}, C., {Meynet}, G., {Maeder}, A., \& {Schaller}, G.
  1993, \aaps, 102, 339

\bibitem[{{Schaller} {et~al.}(1992){Schaller}, {Schaerer}, {Meynet}, \&
  {Maeder}}]{Schaller92}
{Schaller}, G., {Schaerer}, D., {Meynet}, G., \& {Maeder}, A. 1992, \aaps, 96,
  269

\bibitem[{{Spruit}(2008)}]{Spruit08}
{Spruit}, H.~C. 2008, in American Institute of Physics Conference Series, Vol.
  983, 40 Years of Pulsars: Millisecond Pulsars, Magnetars and More, ed.
  C.~{Bassa}, Z.~{Wang}, A.~{Cumming}, \& V.~M. {Kaspi}, 391--398

\bibitem[{{Testa} {et~al.}(2008){Testa}, {Rea}, {Mignani}, {Israel}, {Perna},
  {Chaty}, {Stella}, {Covino}, {Turolla}, {Zane}, {Lo Curto}, {Campana},
  {Marconi}, \& {Mereghetti}}]{Testa08}
{Testa}, V., {Rea}, N., {Mignani}, R.~P., {Israel}, G.~L., {Perna}, R.,
  {Chaty}, S., {Stella}, L., {Covino}, S., {Turolla}, R., {Zane}, S., {Lo
  Curto}, G., {Campana}, S., {Marconi}, G., \& {Mereghetti}, S. 2008, \aap,
  482, 607

\bibitem[{{Thompson} \& {Duncan}(1995)}]{T-D95}
{Thompson}, C. \& {Duncan}, R.~C. 1995, \mnras, 275, 255

\bibitem[{{Thompson} {et~al.}(2004){Thompson}, {Chang}, \& {Quataert}}]{TCQ04}
{Thompson}, T.~A., {Chang}, P., \& {Quataert}, E. 2004, \apj, 611, 380

\bibitem[{{Vink} \& {Kuiper}(2006)}]{V-K06}
{Vink}, J. \& {Kuiper}, L. 2006, \mnras, 370, L14

\bibitem[{{Vrba} {et~al.}(2000){Vrba}, {Henden}, {Luginbuhl}, {Guetter},
  {Hartmann}, \& {Klose}}]{Vrba00}
{Vrba}, F.~J., {Henden}, A.~A., {Luginbuhl}, C.~B., {Guetter}, H.~H.,
  {Hartmann}, D.~H., \& {Klose}, S. 2000, \apjl, 533, L17

\bibitem[{{Vrba} {et~al.}(1996){Vrba}, {Luginbuhl}, {Hurley}, {Li}, {Kulkarni},
  {van Kerkwijk}, {Hartmann}, {Campusano}, {Graham}, {Clowes}, {Kouveliotou},
  {Probst}, {Gatley}, {Merrill}, {Joyce}, {Mendez}, {Smith}, \&
  {Schultz}}]{Vrba96}
{Vrba}, F.~J., {Luginbuhl}, C.~B., {Hurley}, K.~C., {Li}, P., {Kulkarni},
  S.~R., {van Kerkwijk}, M.~H., {Hartmann}, D.~H., {Campusano}, L.~E.,
  {Graham}, M.~J., {Clowes}, R.~G., {Kouveliotou}, C., {Probst}, R., {Gatley},
  I., {Merrill}, M., {Joyce}, R., {Mendez}, R., {Smith}, I., \& {Schultz}, A.
  1996, \apj, 468, 225

\bibitem[{{Wachter} {et~al.}(2008){Wachter}, {Ramirez-Ruiz}, {Dwarkadas},
  {Kouveliotou}, {Granot}, {Patel}, \& {Figer}}]{Wachter08}
{Wachter}, S., {Ramirez-Ruiz}, E., {Dwarkadas}, V.~V., {Kouveliotou}, C.,
  {Granot}, J., {Patel}, S.~K., \& {Figer}, D. 2008, \nat, 453, 626

\bibitem[{{Wallace} \& {Hinkle}(1996)}]{W-H96arct}
{Wallace}, L. \& {Hinkle}, K. 1996, \apjs, 103, 235

\end{thebibliography}

\end{document}